\documentclass[prd,twocolumn,floatfix,preprintnumbers,showpacs]{revtex4}
\usepackage{graphicx}
\usepackage{dcolumn}
\usepackage{bm}
\usepackage{color}
\usepackage{hyperref}
\usepackage{amsmath} 

\def\lsim{\mathrel{\mathop
  {\hbox{\lower0.5ex\hbox{$\sim$}\kern-0.8em\lower-0.7ex\hbox{$<$}}}}}
\def\gsim{\mathrel{\mathop
  {\hbox{\lower0.5ex\hbox{$\sim$}\kern-0.8em\lower-0.7ex\hbox{$>$}}}}}

\begin{document}

\preprint{LAPTH-1213/07}

\title{Constraining neutrino masses with the 
ISW-galaxy correlation function}

\author{Julien Lesgourgues}
\email{julien.lesgourgues@lapp.in2p3.fr}
\affiliation{LAPTH, Universit\'e de Savoie \& CNRS, BP110, F-64941 Annecy-le-vieux Cedex, France} 

\author{Enrique Gazta\~naga}
\email{gazta@ieec.uab.es}
\affiliation{Institut de Ci\`encies de l'Espai, CSIC/IEEC, Campus UAB,
F. de Ci\`encies, Torre C5 par-2,  Barcelona 08193, Spain}

\author{Wessel Valkenburg}
\email{wessel.valkenburg@lapp.in2p3.fr}
\affiliation{LAPTH, Universit\'e de Savoie \& CNRS, BP110, F-64941 Annecy-le-vieux Cedex, France} 

\date{March 5, 2008}

\begin{abstract}
Temperature anisotropies in the Cosmic Microwave Background (CMB) are
affected by the late Integrated Sachs-Wolfe (lISW) effect caused by
any time-variation of the gravitational potential on linear scales.
Dark energy is not the only source of lISW, since massive neutrinos
induce a small decay of the potential on small scales during both
matter and dark energy domination. In this work, we study the prospect
of using the cross-correlation between CMB and galaxy density maps as
a tool for constraining the neutrino mass. On the one hand massive
neutrinos reduce the cross-correlation spectrum because free-streaming
slows down structure formation; on the other hand, they enhance it
through their change in the effective linear growth. We show that in
the observable range of scales and redshifts, the first effect
dominates, but the second one is not negligible. We carry out an error
forecast analysis by fitting some mock data inspired by the Planck
satellite, Dark Energy Survey (DES) and Large Synoptic Survey
Telescope (LSST).  The inclusion of the cross-correlation data from
Planck and LSST increases the sensitivity to the neutrino mass $m_{\nu}$ by 38\% (and to
the dark energy equation of state $w$ by 83\%) with respect to Planck alone. The
correlation between Planck and DES brings a far less significant
improvement. This method is not potentially as good for detecting
$m_{\nu}$ as the measurement of galaxy, cluster or cosmic shear power
spectra, but since it is independent and affected by different
systematics, it remains potentially interesting if the total neutrino
mass is  of the order of 0.2~eV; if instead it is close to the lower bound
from atmospheric oscillations, $m_{\nu} \sim 0.05$~eV, we do not
expect the ISW-galaxy correlation to be ever sensitive to $m_{\nu}$.
\end{abstract}

\maketitle

\section{Introduction}
As photons pass through a changing gravitational potential well, they
experience a redshift or a blueshift, depending on whether the
well grows or decays respectively. Cosmic microwave background (CMB)
photons can experience such variations between the time of last
scattering and their detection now. This effect was first described by
Sachs and Wolfe in 1967~\cite{Sachs:1967er}, and hence is dubbed the
integrated Sachs-Wolfe effect (ISW). During a Cold Dark Matter (CDM)
and/or baryon dominated era, the gravitational potential distribution
remains frozen, and the ISW effect has no net effect on the blackbody
temperature of CMB photons. This property is crucially related to the
fact that non-relativistic matter (like CDM and baryons) has a
vanishing sound speed, and experiences gravitational clustering on
all sub-Hubble scales after photon decoupling, as
described by the Poisson equation. In such a situation, the universal
expansion and the gravitational contraction compensate each other in
such a way as to maintain a static gravitational potential. However, when
the expansion rate is affected by any type of matter with a
non-vanishing sound speed, e.g. during Dark Energy (DE) domination, the
gravitational perturbations decay and the cosmic photon fluid
experiences a blue shift, acquiring extra temperature perturbations
related to the intervening pattern of matter perturbations. It was first
proposed by Crittenden and Turok in 1995~\cite{Crittenden:1995ak} to
cross correlate maps of temperature perturbations in the CMB with
those of matter overdensities in large scale structures (LSS), in
order to measure a possible acceleration of the universe's expansion.
However, the CMB and LSS data available at that time were not good
enough for such an ambitious goal, and the first strong indication of
a positive acceleration came in 1998 from the side of type-Ia
supernovae~\cite{Perlmutter:1998np,Riess:1998cb}.  Analyses of the
first (2003) and second (2006) data releases of the Wilkinson
Microwave Anisotropy Probe (WMAP)~\cite{Spergel:2003cb,Spergel:2006hy}
were the first to indicate the existence of Dark Energy independent of
acceleration, by means of the location of the second peak in the CMB
power spectrum.  Simultaneously, a number of interesting papers
presented the first detections of the ISW effect by cross-correlating WMAP
anisotropy maps with various LSS data
sets
~\cite{Fosalba:2003ix,Boughn:2003yz,Fosalba:2003iy,Afshordi:2003xu,
Padmanabhan:2004,Cabre:2006qm,Cabre:2006uj,Giannantonio:2006du,McEwen:2007},
now able to give an independent measure for the acceleration of the
expansion of the universe.

The domination of Dark Energy is not the only source of gravitational
potential evolution and of a net ISW effect. On small cosmological
scales, as soon as matter perturbations exceed the linear regime,
gravitational perturbations start to grow and to redshift CMB
photons. This effect, called the Rees-Sciama effect, has not been
significantly detected until now~\cite{Puchades:2006gs}. CMB photons
can also be scattered by gravitational lensing \cite{SeljakZaldarriaga:2000}
and by the Sunyaev-Zeldovich (SZ)  effect \cite{SZ:1969} 
(see \cite{Fosalba:2003iy,Afshordi:2003xu,Afshordi:2007} for detections
in CMB-LSS cross-correlation analysis). An other party
expected to affect the evolution of gravitational perturbations --at
least by a small amount-- is the background of massive neutrinos. Over
thirty years ago massive neutrinos were proposed as a Hot Dark Matter
(HDM) candidate, and later ruled out as the dominant dark component,
since HDM tends to wash out small scale overdensities during structure
formation~\cite{Primack:2001ib}.  Observed neutrino oscillations
however constrain neutrinos to have a
mass~\cite{Maltoni:2004ei,Fogli:2005cq}. In addition, the presence of
a Cosmic Neutrino Background (CNB) is strongly suggested on the one
hand by the abundance of light elements produced during primordial
nucleosynthesis~\cite{Cuoco:2003cu,Steigman:2005uz,Mangano:2006ur}, 
and on the other hand by CMB
anisotropies~\cite{Crotty:2003th,Pierpaoli:2003kw,Barger:2003zg,Trotta:2004ty,Hannestad:2005jj,Hannestad:2006mi,Ichikawa:2006vm,Ichikawa:2006vm,deBernardis:2007bu,Hamann:2007pi}. Therefore,
a small fraction of HDM is expected to coexist with the dominant CDM
component.
%
On small cosmological scales (for instance, cluster scale),
the free-streaming of massive neutrinos should induce a slow decay of
gravitational and matter perturbations~\cite{Bond:1980ha}, acting
during both matter and Dark Energy domination. 
This effect depends on the total neutrino
mass summed over all neutrino families, $m_{\nu}=\sum_i m_i$, unlike
laboratory experiments based on tritium decay or neutrinoless
double-beta decay, which probe different combinations: hence, a
cosmological determination of the total neutrino mass would bring
complementary information to the scheduled particle physics experiments~\cite{Hannestad:2006zg,Lesgourgues:2006nd}. 
The free streaming of massive neutrinos has not yet
been detected~\cite{Hannestad:2007tu}, but there are good prospects to do so in the future,
since the smallest total neutrino mass allowed by data on atmospheric neutrino oscillations 
($m_{\nu} \geq \sqrt{\Delta m^2_{\rm atm}} \sim 0.05$~eV)
implies at least a 5\% suppression in the matter/gravitational
small-scale power spectrum~\cite{Hannestad:2006zg,Lesgourgues:2006nd}.
A positive detection --even in the case of minimal mass-- could follow
from the analysis of future galaxy/cluster redshift surveys~\cite{Lesgourgues:2004ps,Wang:2005vr,Hannestad:2007cp}, weak
lensing surveys~\cite{Song:2003gg,Hannestad:2006as}, Lyman-$\alpha$ forest analysis, cluster counts~\cite{Wang:2005vr},
etc. The goal of measuring the neutrino mass from cosmology is very
ambitious since each of these methods suffers from its own source of
systematics (bias issues, modeling of non-linear clustering,
...). Therefore, a robust detection could only be achieved by
comparing the results from various types of experiments.

The goal of this work is to describe a possible cosmological
determination of the absolute neutrino mass scale through the ISW
effect induced by neutrino free-streaming on CMB temperature maps,
using as an observable the cross-correlation function of
galaxy-temperature maps. This possibility was investigated previously
by Ichikawa and Takahashi \cite{Ichikawa:2005hi} (and suggested again
recently in \cite{Kiakotou:2007pz}).
As neutrinos slow down the growth of structure, we expect the
blueshift caused by an accelerated expansion to be more pronounced if
neutrinos have a larger mass. On the other hand, the distribution of
matter inducing the late ISW effect is smoother in case of
free-streaming by massive neutrinos. These two antagonist effects
should in principle induce some mass-dependent variations in the
galaxy-temperature cross-correlation function.

In section~\ref{sec:theory} of this paper we give an outline of the
theory of the ISW-effect in the presence of a neutrino mass. In
section~\ref{sec:mock}, we use some mock data with properties inspired
from the Planck satellite, Dark Energy Survey (DES) and Large Synoptic
Survey Telescope (LSST) in order to show the potential impact of this
method in the future.

\section{The galaxy-ISW correlation in the presence of neutrino mass}
\label{sec:theory}

\subsection{Definitions}

The observed galaxy overdensity $\delta_G$ in a given direction
$\hat n$ is defined as
\begin{align}
  \delta_G(\hat n)=\int dz\, b(z)\phi_G(z)\delta_m(\hat n, z),
\end{align}
where $z$ denotes redshift, $b(z)$ is the redshift dependent bias
function relating the observed galaxy overdensity to the total matter 
overdensity, and $\phi_G(z)$ is the galaxy selection
function which can be chosen such that only galaxies within a certain
range of redshift are considered.

The observed CMB temperature map
\begin{align}
\Delta_T(\hat n)\equiv
\frac{T(\hat n)-T_0}{T_0}
\end{align}
results from various contributions, classified as primary or secondary
anisotropies. By definition, secondary anisotropies are induced after
photon decoupling and can be correlated to some extent with the
surrounding large scale structure.  The ISW component is one of these
terms, and can be obtained by integrating the scalar metric
perturbations (or just the Newtonian gravitational potential on
sub-Hubble scales) along each line-of-sight between the last
scattering surface and the observer.  If the gravitational potential
is written as a function of direction $\hat n$ and redshift $z$, the
ISW term reads
\begin{align}
  \Delta_T^{ISW}(\hat n)= -2\int_{0}^{z_{\rm dec}} dz\, \frac{d \Phi}{dz}(\hat n, z).
\end{align}
where ${z_{\rm dec}} $ is the redshift at decouplig.
Immediately after decoupling and before full matter domination, the
gravitational potential does vary with time: this is known as the
early ISW (eISW) effect, in contrast with the late ISW (lISW) in which
we are presently interested. The two maps $\Delta_T^{eISW}$,
$\Delta_T^{lISW}$ can be computed separately by cutting the above
integral in two pieces at some intermediate redshift $z_*$ chosen
during full matter domination, when the gravitational potential is
static. Note that in presence of massive neutrinos, the potential is
never really static on small scales, so the quantity $\Delta_T^{lISW}$
might not be uniquely defined. Anyway, this question is not relevant
in practice.  The observable quantity is not the late ISW
auto-correlation function $\langle \Delta_T^{lISW}(\hat n)
\Delta_T^{lISW}(\hat n') \rangle$, but only its cross-correlation with
a given survey $\langle \Delta_T^{lISW}(\hat n) \delta_G(\hat n')
\rangle$. Then, the redshift distribution $\phi_G(z)$ selects the
range in which the ISW effect is being probed, and the choice of $z_*$
becomes irrelevant provided that $z_*$ remains larger than the
redshift of all objects in the survey: $\phi_G(z_*) \simeq 0$.


Assuming that the galaxy-temperature cross-correlation function arises solely
from the late ISW effect (i.e., assuming that other secondary anisotropies
potentially correlated with LSS can be separated or have a
negligible amplitude, which is a good assumption on the scales
considered hereafter), we can relate the galaxy-temperature
correlation multipoles to the real-space correlation function
$\langle \Delta_T^{lISW}(\hat{n}) \delta_G(\hat n') \rangle$. In the Limber
approximation (see Appendix), one gets
\begin{align}
C_l^{TG} \! 
=& \frac{3 \Omega_m H_0^2}{(l \! + \! 1/2)^2} \label{eq:C_l_TG} \\ 
&\times \int_0^{z_*} \!\! dz\, b(z) \phi_G(z) H(z)
a(z) \left[ \partial_z \frac{P(k,z)}{a(z)^2} \right]_{k=\frac{l+1/2}{r(z)}}
, \nonumber
\end{align}
where $r(z)$ is the conformal distance up to redshift $z$, 
$H_0=100h$~km/s/Mpc is the Hubble parameter today, and the
matter power spectrum is defined as $\langle
\delta_m(\vec{k},z)\delta_m(\vec{k}',z) \rangle \equiv P(k,z) \, \,
\delta^3(\vec k-\vec k')$. Note that we used the Poisson equation in
flat space in order to relate the gravitational potential $\Phi$ to
the matter overdensity $\delta_m$, and assumed
$a(0)=1$ by convention. Finally, the multipoles $C_l^{TG}$ define
the angular correlation function in a Legendre polynomial basis
($p_l$),
\begin{align}
  w^{TG}(\theta)=\sum_l \frac{2l+1}{4\pi}p_l(\cos\theta)C^{TG}_l.
\label{eq:w_TG}
\end{align}
Eq.(\ref{eq:C_l_TG}) is often written in a form which assumes
that the matter power spectrum is a separable function of wavenumber
and redshift. This applies to the case of a (flat) $\Lambda$CDM
universe, for which one can write
\begin{align}
P(k,z)= D(\Lambda;z)^2 a(z)^2 P(k,0)
\label{eq:D_Lambda}
\end{align}
with $\partial_z D=0$ during full matter domination and $\partial_z
D>0$ during $\Lambda$ domination. Figure~\ref{fig:D} (left) shows the
evolution of $D$ as a function of $z$ for
$\Omega_{\Lambda}=1-\Omega_m=0.69$.  In the case of time-varying Dark
Energy, the situation is qualitatively similar, and $D$ just depends
on more free parameters than $\Lambda$. In the rest of this paper, we
will just write this function as $D(z)$ for concision.

\subsection{Effect of neutrino masses}

\begin{figure}[t]
\includegraphics[width=4.25cm]{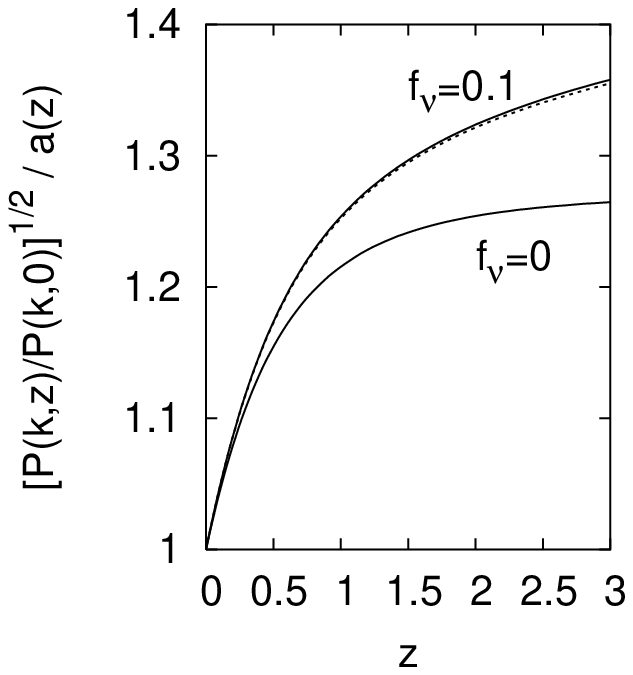}
\includegraphics[width=4.25cm]{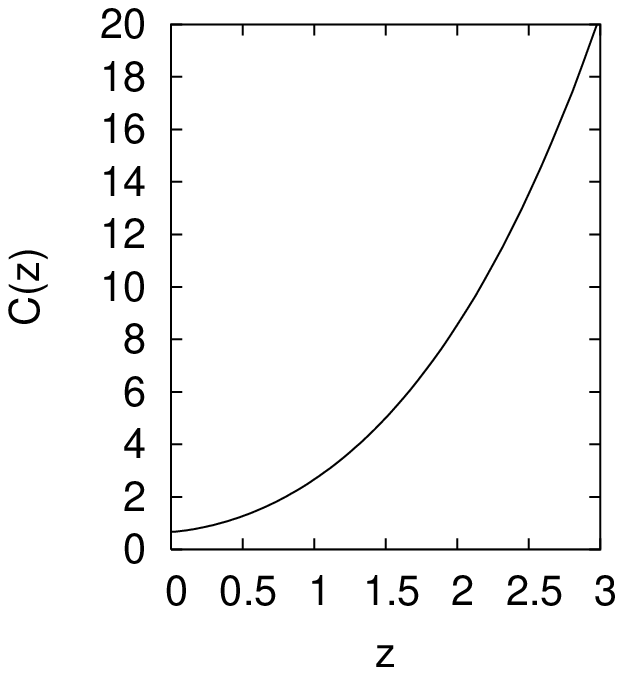}
\caption{{\it (Left)} ~Redshift evolution of the small-scale linear
growth factor, defined here as $[P(k,z)/P(k,0)]^{1/2} / a(z)$ for 
$k \sim 10 \, h {\rm Mpc}^{-1}$, and obtained numerically with {\sc camb}
for $\Omega_{\Lambda}=0.69$.
The lower curve corresponds to $f_{\nu}=0$ and is exactly equal to
the quantity $D(z)$ defined in Eq.(\ref{eq:D_Lambda}). The upper,
solid curve corresponds to $f_{\nu}=0.1$, and is well approximated
by the dotted curve, which corresponds to the first of Eqs.(\ref{eq:D_fnu}).
{\it (Right)} ~The function $C(z)$, defined in Eqs.(\ref{eq:C}), computed here
for $\Omega_{\Lambda}=0.69$. Roughly speaking, the effect of neutrino masses
on $C_l^{TG}$ changes of sign when this function crosses eighth.}\label{fig:D}
\end{figure}

In models with massive neutrinos, the spectrum is not a separable
function anymore (in other words, the linear growth factor is
scale-dependent), and Eq.(\ref{eq:C_l_TG}) cannot be further
simplified. However, in order to make analytical estimates of the
impact of neutrino masses on $C_l^{TG}$, it is possible to use some
approximate solutions valid only on the largest and smallest
wavelength~(see \cite{Lesgourgues:2006nd} and \cite{Kiakotou:2007pz}
for more details). First, for wavelengths larger than the maximum
value of the neutrino free-streaming scale, reached at the time of the
transition to the non-relativistic regime, the power spectrum
$P^{f_{\nu}}$ is completely unaffected by neutrino masses, and
identical to that in a massless neutrino model with the same
cosmological parameters (in particular, the same $\Omega_m$ and $h$)
noted as $P^0$:
\begin{align}
\forall k < k_{\rm nr},&  &
P^{f_{\nu}}(k,z)&=[D(z) a(z)]^2 P^{f_{\nu}}(k,0) \nonumber \\
{\rm with}& & P^{f_{\nu}}(k,0)&=P^0(k,0)~.
\end{align}
On the other hand, for wavelengths smaller than the the free-streaming scale
today, both the linear growth factor and the amplitude today are
affected by neutrino masses, approximately like:
\begin{align}
\forall k > k_{\rm fs},&  &
P^{f_{\nu}}(k,z)\simeq & \, [D(z) a(z)]^{2-\frac{6}{5} f_{\nu}} 
P^{f_{\nu}}(k,0) \nonumber \\
{\rm with}& & P^{f_{\nu}}(k,0) \simeq & \, [1-8 f_{\nu}] P^0(k,0)~,
\label{eq:D_fnu}
\end{align}
where $f_{\nu}=\Omega_{\nu}/\Omega_m$ stands for the neutrino density
today relative to the {\it total} matter density (so $\Omega_m$
includes baryons, hot and cold dark matter). Here $D(z)$ is always the
same function, computed either for $f_{\nu} \neq 0$ on large scales,
or for $f_{\nu} = 0$ on any scale, with a common value of
$\Omega_{\Lambda}$ (or of Dark Energy parameters).  The first
approximation in Eqs.~(\ref{eq:D_fnu}) is very accurate, as shown in
Fig.~\ref{fig:D} (left) where we compare the precise linear growth factor
obtained numerically with the above solution.  The second
approximation is poorer, but more accurate ones can be found e.g. in
Refs.~\cite{Lesgourgues:2006nd,Kiakotou:2007pz}.

Assuming that the galaxy selection function is very peaked around a
median redshift $z_m$, the multipole $C_l^{GT}$ probes mainly
fluctuations around the scale $k \sim l/r(z_m)$. If $l$ is larger than
$k_{\rm fs} \, r(z_m)$, $C_l^{GT}$ is affected by neutrino masses through the
term between brackets in Eq.~(\ref{eq:C_l_TG}). Using
Eqs.~(\ref{eq:D_fnu}), this term varies with $f_{\nu}$ like:
\begin{align}
\partial_z \frac{P^{f_{\nu}}(k,z)}{a(z)^2} &\simeq
\left[ \left(1 + C(z) f_{\nu} \right) \left( 1-8f_{\nu} \right) \right]
\partial_z \frac{P^{0}(k,z)}{a(z)^2}~,
\label{eq:C} \\
{\rm with }~~~~~~ 
& C(z)=\frac{3}{5} \left(\frac{1}{1+z}\frac{D}{D'} - 1 \right)~. \nonumber
\end{align}
For a typical Dark Energy model,
the density fraction $\Omega_{\rm DE}$
becomes negligible for $z>2$
\footnote{in the case of ``early Dark Energy'' models, this statement
can only be marginally true (see e.g. in
\cite{Doran:2006kp,Doran:2005ep})}, and hence the ratio $D'/D$ is
tiny. So, at hight redshift, the net effect of the neutrino mass is to
increase the integrand in $C_l^{TG}$ like:
\begin{align}
\partial_z \frac{P^{f_{\nu}}(k,z)}{a(z)^2} \simeq 
\left[ \frac{3}{5} \frac{f_{\nu} (1 - 8 f_{\nu})}{1+z} \frac{D}{D'} \right]
\, \, \partial_z \frac{P^{0}(k,z)}{a(z)^2}~.
\end{align}
This just reflects the fact that at high redshift, the ISW effect
would be null on all scales for $f_{\nu}=0$, while for $f_{\nu}>0$ it
is still active on small scales. However, for $z<2$, $D'/D$ becomes
larger, and for typical values of $\Omega_{\rm DE}\sim 0.7$ 
there is always a redshift below which $C(z)$ is smaller than eight.
Then, the term between brackets in Eq.~(\ref{eq:C}) is smaller than one,
and the net effect of neutrino masses is to decrease
$\partial_z [P / a^2]$. In Fig.~\ref{fig:D} (right), we plot
the function $C(z)$ in the case of a cosmological constant with
$\Omega_{\Lambda}=0.69$. We see that $C \sim 8$ for $z \sim 2$;
so, around this redshift and for $l > k_{\rm fs} \, r(z_m)$, 
the net effect of neutrino masses on
$C_l^{GT}$ changes of sign.

In summary, if $z_m$ is small, the expected effect of neutrino masses
on the cross-correlation multipoles $C_l^{GT}$ consists in a step-like
suppression at large $l$'s, qualitatively similar to that observed in
the galaxy auto-correlation multipoles $C_l^{GG}$. However the
suppression factor is smaller, since the lack of power in the matter
power spectrum caused by neutrino free-streaming is balanced by the
excess of ISW effect due to the behavior of the linear growth factor
in presence of massive neutrinos. When $z_m$ increases, the boost
related to the ISW effect is seen more clearly, and ultimately,
when $z_m$ is chosen before dark energy domination, the net effect of
neutrino masses is to increase $C_l^{GT}$ at large $l$.

In order to check and quantify these effects, we computed the
cross-correlation multipoles $C_l^{TG}$ (and also for comparison the
auto-correlation multipoles $C_l^{GG}$) for two different cosmological
models, sharing the same parameters $\Omega_b=0.053$, $\Omega_m=0.31$,
$\Omega_{\Lambda}=0.69$, $h=0.65$, $A \equiv \ln [ 10^{10} k^3 {\cal
R}^2 ]_{k=0.01/{\rm Mpc}}=3.16$, $n_s=0.95$, but with two different
values of the neutrino density fraction $f_\nu=\Omega_{\nu}/\Omega_m$,
equal to $0$ or $0.1$ (corresponding to three neutrino species sharing the
same mass $m_{\nu}=0$ or $m_{\nu}\simeq0.41$~eV).  We adopted a galaxy
selection function of the form
\begin{align}
  \phi_G(z)=\frac32\frac{z^2}{z_0^3} \exp\left[{-\left(\frac{z}{z_0}\right)^{\frac32}}\right]~,
\end{align}
peaking near the median redshift $z_m \equiv 1.4 z_0$. For
illustrative purposes, we choose the four values $z_m=0.1, 1, 2, 3$, 
although in practice it would be very challenging to map
$\delta_G(\hat{n})$ for $z\geq 2$: presently, available data with
a reasonable signal-to-noise ratio range only from
$z \sim 0.1$ to $z \sim 1.5$.

\begin{figure}[t]
\includegraphics[width=8.5cm]{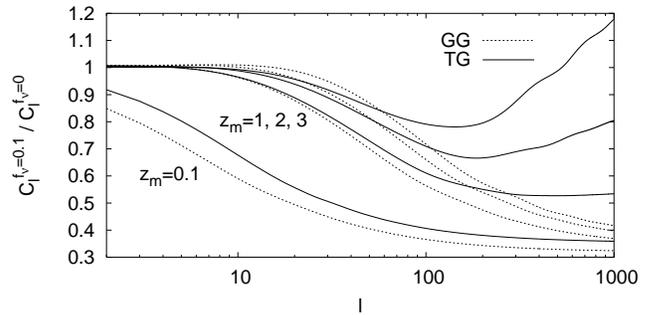}
\caption{Ratio of the cross-correlation multipoles $C_l^{TG}$ and
auto-correlation multipoles $C_l^{GG}$ obtained for two cosmological
models with neutrino density fractions equal to $f_{\nu}=0.1$ or 0,
and the same value of other cosmological parameters (see the text for details).
}\label{fig:cl_ratio}
\end{figure}

In Fig.~\ref{fig:cl_ratio} we plot the ratio of the multipoles
$C_l^{TG}$ in the two models, compared with the same ratio for
$C_l^{GG}$. The free-streaming of massive neutrinos is responsible for
the step-like suppression of $C_l^{GG}$, like in the power spectrum
$P(k)$. The value of $z_m$ controls the angle under which the
free-streaming scale is seen in the map $\delta_G(\hat{n})$, and hence
the scale at which the suppression occurs in multipole space. As
expected from the previous discussion, the neutrino mass effect on
$C_l^{TG}$ is similar to that on $C_l^{GG}$ for small $z_m<1$,
although the suppression factor is slightly smaller, due to the excess
of ISW effect in presence of massive neutrinos. For $z_m \geq 1$, the
amplification effect due to this excess has a clear and distinct
signature at $l\geq100$, and for $z_m \sim 2$ the ratio displayed in
Fig.~(\ref{fig:cl_ratio}) has a dip around $l \sim 150$. Unfortunatly,
we will see in Sec.~\ref{sec:detectability} that for $l \geq 100$
this effect is masked by primary CMB anisotropies, which play the role of
white noise for the present purpose.

\begin{figure}[t]
\includegraphics[width=8.5cm]{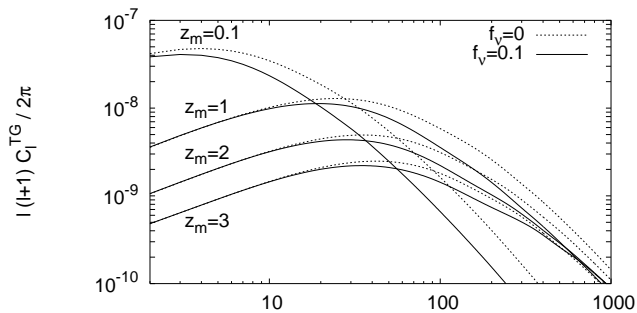}
\caption{Dimensionless cross-correlation spectrum in multipole space,
$l(l+1)C_l^{TG}/2 \pi$, for the same cosmological models as in
Fig.~\ref{fig:cl_ratio} (i.e., with three neutrino species sharing the
same mass $m_{\nu}=0$ or $m_{\nu}\simeq0.41$~eV).}\label{fig:cl}
\end{figure}

In Fig.~\ref{fig:cl}, we plot directly the mutipoles $C_l^{TG}$ for
the same two models. The effect of neutrino masses is clearly visible
for all $l > 2$ at $z_m=0.1$, while for $z_m \geq 1$ it is necessary
to reach $l \geq 20$ in order to see a difference (since the maximum
free-streaming scale is seen under a smaller angle at higher
redshift). Remembering that the effect of neutrino masses on large
$l$'s can be split in two contributions, a matter power suppression
and an excess of ISW, it is clear from the previous discussion that
the latter effect contributes at all redshifts, but its most obvious
manifestation is the fact that $C_l^{TG}$ increases with $f_{\nu}$ for
large $l$'s. However, we will see in Sec.~\ref{sec:detectability} that
only the region with $l \leq 100$ can be probed by observations: then,
the neutrino-induced ISW effect is significant, but smaller that the 
opposite suppression effect.

\begin{figure}[t]
\includegraphics[width=8.5cm]{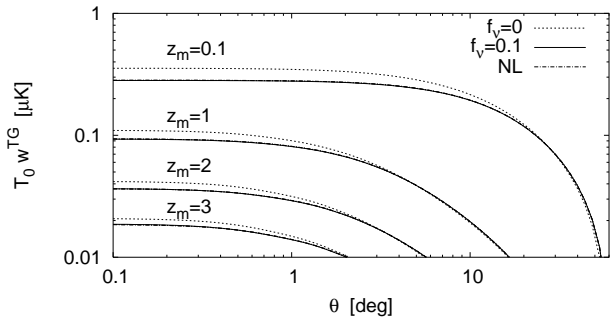}
\caption{Angular cross-correlation function, multiplied by the average
CMB temperature and displayed in units of micro-Kelvins, for the same
cosmological models as in Figs.~\ref{fig:cl_ratio},\ref{fig:cl}. We
also plot the same functions including non-linear (NL) corrections to
the matter power spectrum: they are indistinguishable from the linear
ones.}\label{fig:w}
\end{figure}

In Fig.~\ref{fig:w}, we plot the corresponding angular correlation
functions $w^{TG}(\theta)$. In this representation, the fine-structure
of the high-$l$ multipole spectrum is by construction averaged out,
and it is not possible to see an amplification at high $z_m$ and small
$\theta$. The suppression caused by neutrino masses is visible for
$z_m=0.1$ at $\theta \leq 15^o$, and for $z_m \geq 1$ at 
$\theta \leq 2^o$.

In all these plots, we used only the linear perturbation
theory. Doing so, the angular cross-correlation functions depend on the
matter power spectrum inside the linear regime. To prove it, we
compute again $w^{TG}(\theta)$ from the non-linear power spectrum
obtained by applying {\sc halofit} corrections \cite{Smith:2002dz} to
the linear one. The result, superimposed in Fig.~\ref{fig:w}, is
indistinguishable from that of linear theory. This shows that
non-linear effects on the evolution of matter perturbations has much
less impact than that of adding a neutrino mass. This is also true
for the multipoles $C_l^{TG}$, excepted for the smallest redshifts and highest
$l$'s (for $z_m=0.1$, non-linear effects become important for $l>100$).

\subsection{Detectability}
\label{sec:detectability}

For a set of full-sky CMB and LSS experiments measuring the
temperature multipoles $a_{lm}^T$ (resp. galaxy-density multipoles
$a_{lm}^G$) with a noise spectrum $N_l^T$ (resp. $N_l^G$), the
cross-correlation spectrum $C_l^{TG}$ can be reconstructed from the
estimator
\begin{align}
\tilde{C}_l^{TG}=\frac{\sum_{m=-l}^l a_{lm}^{T*} a_{lm}^G}{2l+1}
\end{align}
with a variance $\sigma_l^{TG}$ given by
\begin{align}
(\sigma_l^{TG})^2
= \frac{(C_l^{TG})^2+(C_l^{TT} + N_l^{TT})(C_l^{GG} + N_l^{GG})}{2l+1}~.
\label{sigmaTG1}
\end{align}
Note that the estimator is not Gaussian, especially for small $l$'s:
so, $\sigma_l^{TG}$ is only an estimate of the true (asymmetric) error
bar on the reconstructed power spectrum.  If the cross-correlation map
can be reconstructed only inside a fraction $f_{\rm sky}$ of the full
sky, in first approximation $\sigma_l^{TG}$ should be multiplied by
$f_{\rm sky}^{-1/2}$. The variance is further reduced by $\sqrt{\Delta
l}$ in case of data binning with bin width $(\Delta l)$. Note that in this
case the covariance matrix is no longer diagonal, but nevertheless 
using a diagonal matrix under these approximations has been shown 
to work well, compared to the exact
treatment, if we choose an adequately large binning \cite{Cabre.etal:2007}.
In practice,
for the multipole range in which we are interested, the CMB noise
spectrum $N_l^{TT}$ is much smaller than $C_l^{TT} $ for experiments
like WMAP and beyond, and can be safely neglected in the above
expression. For a LSS survey consisting in a catalogue of discrete
objects (galaxies, clusters, etc.), the noise spectrum is usually
dominated by the shot noise contribution $N_l^{GG}\simeq 1/\bar{N}$,
where $\bar{N}$ represents the mean number of objects per
steradian. The largest ongoing/future surveys (e.g. SDSS) should reach
typically the order of $10^8$ or even $10^9$.

\begin{figure*}[t]
\includegraphics[height=4cm]{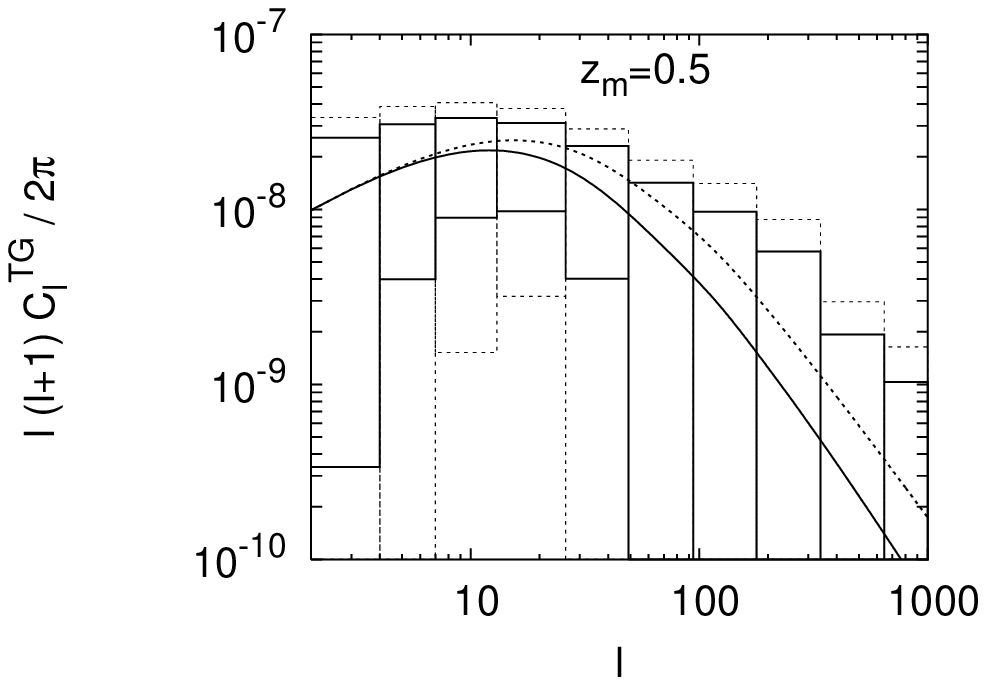}
\hspace{-0.8cm}
\includegraphics[height=4cm]{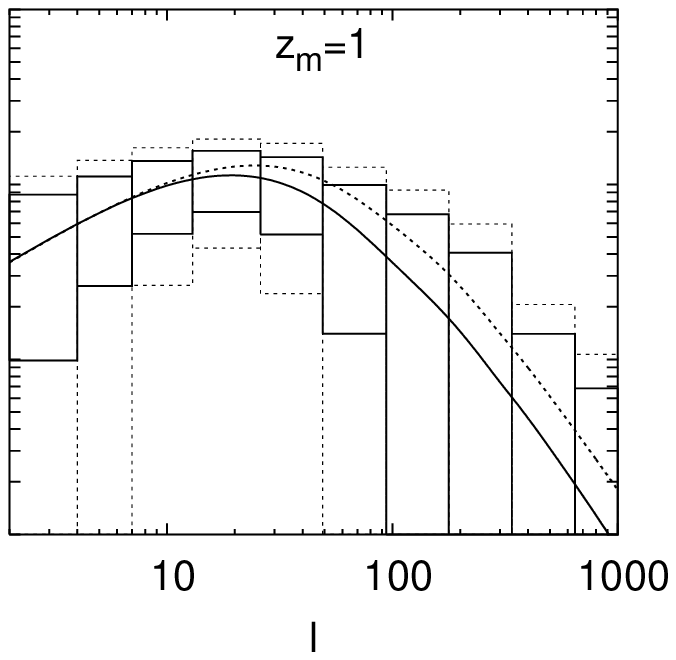}
\hspace{-0.8cm}
\includegraphics[height=4cm]{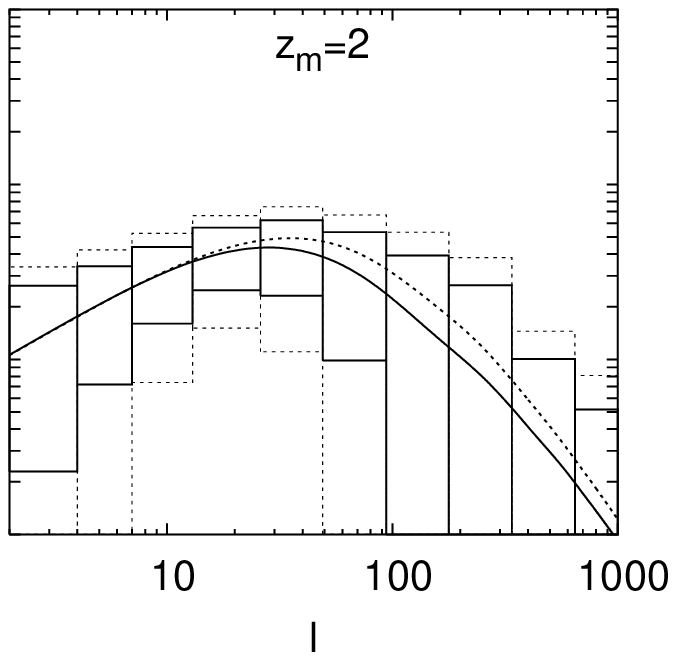}
\hspace{-0.8cm}
\includegraphics[height=4cm]{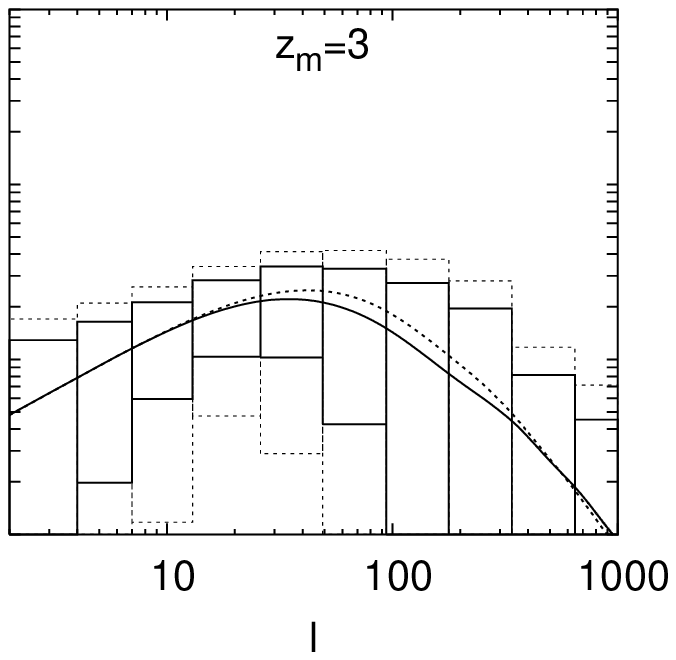}
\caption{68\% error forecast on the power spectrum $C_l^{TG}$, which
is displayed for the same two cosmological models as in
Figures~\ref{fig:cl_ratio}, ~\ref{fig:cl} (with $f_{\nu} = 0$ or
0.1). The smallest error boxes assumes a LSS survey with sky coverage
$f_{\rm sky}=0.65$ and surface density $\bar{N}=10^9 {\rm st}^{-1}$ in
each redshift bin. The binning in multipole space can be read from the
width of each box. The largest error boxes correspond to $f_{\rm sky}
0.25$ and $\bar{N} = 7\times10^8 {\rm st}^{-1}$.  }\label{fig:bins}
\end{figure*}

\begin{figure}[t]
\includegraphics[height=4cm]{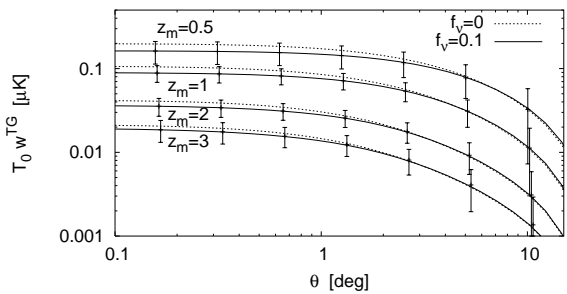}
\caption{68\% error forecast on the angular correlation function
$w^{TG}$, which is displayed for the same two cosmological models as
in Figures~\ref{fig:cl_ratio}, ~\ref{fig:cl} (with $f_{\nu} = 0$ or
0.1). The error bars assume a LSS survey with sky coverage $f_{\rm
sky}=0.65$ and surface density $\bar{N}=10^9 {\rm st}^{-1}$ in each
redshift bin. The spacing between each error bar reflects the
binning width chosen in angular space.}\label{fig:bins_w}
\end{figure}

In Fig.~\ref{fig:bins}, we show the typical errorbar that could be expected
from a cross-correlation map with coverage $f_{\rm sky}=0.65$
(corresponding to the usual galactic cut in CMB maps), using an
ambitious LSS survey with surface density $\bar{N}=10^9 {\rm st}^{-1}$
in each redshift bin. We assumed $b(z)\sim 1$ for simplicity. These
assumptions correspond essentially to the best measurement that
could ever be done, since for such a high surface density the variance
of the estimator of a single multipole product $a_{lm}^{T*} a_{lm}^G$
is not affected by instrumental noise, and reduces to
\begin{align}
\sigma_{lm}^{TG} =  C_l^{TG} \sqrt{1+\frac{C_l^{TT} C_l^{GG}}{(C_l^{TG})^2}}~.
\label{sigmaTG2}
\end{align}
This expression can be interpreted as the product of the cosmic
variance term $C_l^{TG}$ times an enhancement factor depending on the
correlation coefficient $(C_l^{TG})^2/(C_l^{TT} C_l^{GG})$. At large
$l$'s, the late ISW contribution to the total temperature anisotropy
becomes vanishingly small, and the primary anisotropy plays the role
of a large noise term, which cannot be removed. In this limit,
the correlation coefficient is much smaller than one, and the variance
$\sigma_{lm}^{TG}$ gets correspondingly enhanced. Fig.~\ref{fig:bins} shows
that the spectrum $C_l^{TG}$ can be reconstructed to some extent only
in the range $l \leq 100$; beyond, one could only derive upper
bounds. Note that the errorbar for each bin is roughly of the same
order of magnitude as the effect of neutrino masses when $f_{\nu}$
varies from 0 to 0.1.  In Fig.~\ref{fig:bins}, we also show the error
degradation when $f_{\rm sky}$ is reduced to 0.25 and $\bar{N}$ to
$7\times10^8 {\rm st}^{-1}$ 
in each redshift bin. Finally, in Fig.~\ref{fig:bins_w}, we plot
the corresponding error bars for $w^{TG}(\theta)$.  Note that the
synthetic error bars for $w^{TG}(\theta)$ are correlated with each
other, unlike those for $C_l^{TG}$.  On small angular scales $\theta
\leq 1$ (where the effect of neutrino masses is maximal) the $1\sigma$
error on $w^{TG}(\theta)$ is of the order of 25\%.  

We conclude from these estimates that the temperature-galaxy
correlation power spectrum $C_l^{TG}$ is potentially sensitive to the
neutrino mass in the observable range $10< l < 100$, as well as the
angular correlation function $w^{TG}(\theta)$ for $\theta < 5^o$ at $z=0.5$ or $\theta <
3^o$ at $z=1$. Unfortunately, the enhancement of the ISW effect due to the
impact of massive neutrinos on the linear growth factor is not
directly visible: it would require precise data at high $l$ and high
redshift, for which the late ISW effect is masked by primordial
anisotropies. The net effect of massive neutrinos on the observable
part of $C_l^{TG}$ and on $w^{TG}(\theta)$ is a suppression, caused by
the usual free-streaming effect. However this effect is non-trivial in
the sense that $C_l^{GG}$ and $C_l^{TG}$ depend on $f_{\nu}$ through
different relations, due to the fact that the ISW term involves a
time-derivative of the gravitational potential while the galaxy
overdensity does not. Hence, the galaxy-temperature correlation
spectrum can bring some information on neutrino masses which is not
already contained in the sole galaxy auto-correlation spectrum.  In
the next section, we will quantify this statement by performing
a parameter extraction from mock data accounting for future
experiments.

\section{An MCMC analysis of mock data}
\label{sec:mock}

For a given data set consisting in various maps (i.e. multipoles
$a_{lm}^X$) covering a fraction
$f_{\rm sky}$ of the full sky and assumed to obey Gaussian statistics, the
likelihood function ${\cal L}$ is often approximated as
\begin{align}
{\cal L} \propto \Pi_l \left\{
({\rm det} \, C_l^{\rm th})^{-1/2} 
\exp \left[ -\frac{1}{2} {\rm Trace} \, C_l^{\rm obs} {C_l^{\rm th}}^{-1} \right]
\right\}^{(2l+1) f_{\rm sky}}~.\label{likelihood}
\end{align}
where $C_l^{\rm
obs}$ is the data covariance matrix defined by $[C_l^{\rm
obs}]_{XY} = \langle a_{lm}^X  a_{lm}^Y \rangle$, and $C_l^{\rm th}$ the assumed
theoretical covariance matrix for a given fit, which contains the sum of each theoretical power spectrum $C_l^{XY \rm th}$
and of the instrumental noise power spectra $N_l^{XY}$, estimated by
modeling the experiment. Of course, the data covariance matrix 
reconstructed from the observed maps is also
composed of signal and noise contributions.
Simulating a future experimental data set amounts in
computing the noise spectra $N_l^{XY}$, given some instrumental
specifications, and generating randomly some observed spectra
$C_l^{XY \rm obs}$, given the theoretical spectra $C_l^{XY \rm fid}$
of the assumed fiducial
model and the noise spectra $N_l^{XY}$.
However, for the purpose of error forecast, it is sufficient to
replace simply $C_l^{XY \rm obs}$ by the sum
$C_l^{XY \rm fid}+N_l^{XY}$: this just amounts
in averaging over many possible mock data sets for the same model, and does
not change the reconstructed error on model parameters~\cite{Perotto:2006rj}.

For instance,
if one wants to estimate future errors for a CMB experiment, the maps to
consider are temperature and $E$-polarization: $X \in \{T, E\}$ (here,
for simplicity, we consider models with no gravitational waves and 
discard $B$-polarization). The covariance matrices then read
\begin{align}
C_l^{\rm obs} &=&
\left(
\begin{matrix}
C_l^{TT \rm fid} \!\! + \! N_l^{TT} & C_l^{TE \rm fid} \\
C_l^{TE \rm fid} & C_l^{EE \rm fid} \!\! + \! N_l^{EE}
\end{matrix}
\right)~,
\\
C_l^{\rm th} &=&
\left(
\begin{matrix}
C_l^{TT \rm th} \!\! + \! N_l^{TT} & C_l^{TE \rm th} \\
C_l^{TE \rm th} & C_l^{EE \rm th} \!\! + \! N_l^{EE}
\end{matrix}
\right)~.
\end{align}

Should one consider the combination of CMB data with a future galaxy
redshift survey decomposed in $N$ maps associated to $N$ redshift bins,
the matrices would become $2+N$ dimensional, with an extra block
\begin{align}
[C_l]_{2+i,2+j} =
C_l^{G_i G_j} + \delta_{ij} N_l^{G_i G_i}~,~~~i = 1,..., N,
\end{align}
 as well as non-diagonal coefficients $[C_l]_{1,2+i} = C_l^{T G_i}$
accounting for the late ISW effect. Note that all non-diagonal
coefficients have no noise term, since the noise contributions in two
different maps are expected to be statistically uncorrelated at least
at first order.

Finally, the option which is most interesting in our context, is to
assume that the galaxy density auto-correlation maps are not known (or
just not considered, because they could be plagued by some systematic
effects), and that CMB data are only combined with the
cross-correlation data, i.e. with $N$ observed power spectra spectra
$C_l^{TG_i \rm obs}$. This is exactly what is
being done in the current literature, in which authors try to get
some new independent bounds on $\Omega_{\Lambda}$ from CMB
plus CMB-LSS cross-correlation data, without employing LSS
auto-correlation maps.  In the approximation of Gaussian-distributed
$C_l^{TG_i}$ with central value $C_l^{TG_i \rm th}$ and
covariance given by
\begin{align}
&[{\rm Cov}_l]_{ij} \equiv
\left\langle 
\left( C_l^{TG_i} - \langle C_l^{TG_i} \rangle \right)   
\left( C_l^{TG_j} - \langle C_l^{TG_j} \rangle \right)   
\right\rangle \nonumber
\\
&=
\frac{C_l^{TG_i \rm th}C_l^{TG_j \rm th} \!\! +(C_l^{TT \rm th} \!\! + N_l^{TT})(C_l^{G_iG_j \rm th} \!\! + \delta_{ij}N_l^{G_iG_i})}{(2l+1)f_{\rm sky}}~,
\end{align}
the likelihood of the cross-correlation data reads
\begin{align}
{\cal L} \propto \Pi_l \,
({\rm det} \, {\rm Cov}_l)^{-1/2}  
\exp \left[ -\frac{1}{2} \sum_{ij}  \Delta_l^i \,
[{\rm Cov}_l]_{ij}^{-1} \Delta_l^j\right] \label{likelihood2}
\end{align}
with $\Delta_l^i \equiv C_l^{TG_i \rm obs}  -C_l^{TG_i \rm th}$.
The total likelihood is then the product of the CMB and
cross-correlation likelihoods.

In this section, we will focus on three ambitious future experiments:
the Planck satellite, to be launched in 2008, which is expected to
make the ultimate measurement of CMB temperature anisotropies,
dominated by cosmic variance rather than noise up to very high $l$;
the Dark Energy Survey (DES); and
the Large Synoptic Survey Telescope (LSST), designed primarily for
a tomographic study of cosmic shear, which would provide as a
byproduct a very deep and wide galaxy redshift survey (close to ideal
for the purpose of measuring the CMB-LSS cross-correlation since
$N_l^{G_iG_i} < C_l^{G_iG_i}$ at least for multipoles $l<100$).
For Planck, we computed the noise-noise spectra, $N_l^{TT}$ and $N_l^{EE}$,
like in Ref.~\cite{Lesgourgues:2005yv}, with nine frequency channels.
For the DES-like survey, we followed Ref.~\cite{Pogosian:2005ez} and
assumed a total number of galaxies of 250 million in a 5000 square
degree area on the sky (or $f_{\rm sky} = 0.13$), with an approximate
1-$\sigma$ error of 0.1 in photometric redshifts,
divided in four redhsift bins with mean redshifts $z_i \in \{ 0.3, 0.6, 1,
1.3\}$, with the same selection functions as in Ref.~\cite{Pogosian:2005ez}.
For LSST, we used the same modeling as in \cite{LoVerde:2006cj}, with
a net galaxy angular number density of 80 per square arcminute and a
coverage of $f_{\rm sky}=0.65$. The galaxies are divided into six
redshift bins with mean redshifts $z_i \in \{ 0.49, 1.14, 1.93,
2.74,3.54, 4.35\}$. For each bin the selection function, estimated
bias $b_i$ and galaxy density $n_i$ are provided in
\cite{LoVerde:2006cj} (Fig.~2, Eq.~(16) and Table I). The noise
spectra $N_l^{G_iG_i}$ are then simply given by $1/n_i$.

We used the public  code {\sc cosmomc} \cite{Lewis:2002ah} to do a
Monte-Carlo Markov Chain (MCMC) analysis, fitting the theoretical
galaxy-temperature correlation to the mock data. For this purpose, we
have written a module which computes the correlation multipoles
following Eq.~(\ref{eq:C_l_TG}) and the likelihood of the mock data
given each model as described above.

\begin{figure}[t]
\includegraphics[width=8cm]{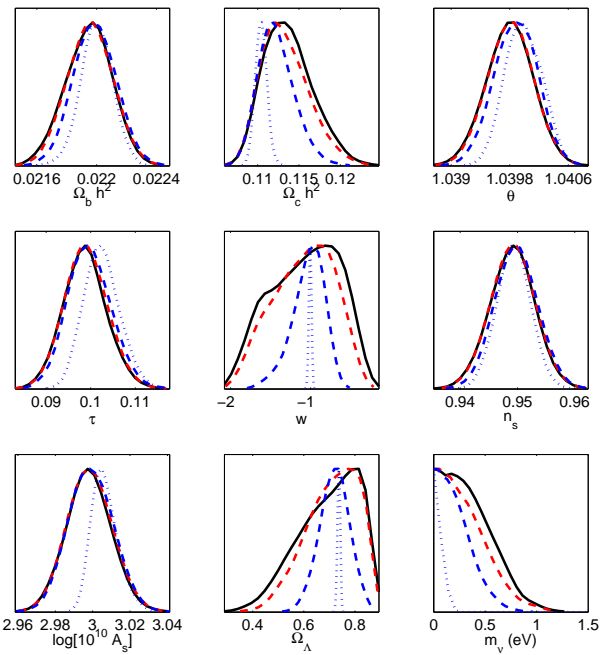}
\caption{Marginalized probability of cosmological parameters obtained
by fitting some mock data mimicking the properties of Planck, DES and LSST.
The solid black curves accounts for CMB only, the red dashed for
CMB+GT with Planck+DES, the blue dashed for
CMB+GT with Planck+LSST, 
and the dotted blue for CMB+GT+GG with Planck+LSST (these combinations are
precisely defined in the text). In each of these cases the cosmological
model consists
in $\Lambda$CDM (six parameter) plus an arbitrary total neutrino
mass $m_{\nu}$ and equation-of-state parameter $w$.
So, only eight of the above nine parameters are independent (as a consequence
the prior for $\Omega_{\Lambda}$ is non-flat). The mock data is based on
a fiducial model with $m_{\nu}=0$ and $w=-1$.}\label{fig:mock_mnuw}
\end{figure}

We then ran our modified version of CosmoMC for a model with eight
parameters: the usual six of minimal $\Lambda$CDM 
(baryon density $\Omega_b h^2$, 
dark matter density $\Omega_{dm} h^2$, 
angular diameter of the sound horizon at last scattering $\theta$,
optical depth to reionization $\tau$,
primordial spectral index $n_s$,
primordial amplitude $\log[10^{10} A_s]$)
plus the total neutrino
mass $m_{\nu}$ and the equation-of-state parameter $w$. Our fiducial
model was close to the WMAP best-fitting model with $m_{\nu}=0$ and
$w=-1$. We considered three possible combinations of data: Planck
alone, Planck plus its cross-correlation with DES or LSST (but no information
on galaxy auto-correlations), and finally Planck plus LSST, using all
information and including the correlation. The probability of each
parameter is displayed in Fig.~\ref{fig:mock_mnuw} for each of these four cases
called respectively CMB (Planck), CMB+GT (Planck+DES or Planck+LSST) 
and CMB+GT+GG (Planck+LSST).

Obviously the combination CMB+GT+GG does a much better job than
CMB+GT for constraining all parameters (and most spectacularly $w$ and
$m_{\nu}$).  This is mainly due to the fact that the GT
cross-correlation is partly screened by primary temperature
anisotropies, while the GG signal does not have such an intrinsic noise
contribution. We even try to repeat the CMB+GT+GG analysis with all
$C_l^{GT_i}$ correlations set to zero, and found no noticeable
difference, showing that most sensitivity comes from GG rather than GT
terms.  However, the comparison between CMB alone and CMB+GT is still
interesting {\it per se}. In fact, we are dealing here with an
idealized situation, but in the future the GG auto-correlation signal
could appear to be plagued by various systematic effects. In this case,
independent information coming from the cross-correlation signal alone
might be a useful piece of evidence in favor of the preferred
model. Also, if the galaxy bias turns out to be very difficult to
estimate with high enough accuracy, one may adopt the point of view
of using the GG signal to measure bias, and the CMB+GT signal to
estimate the best-fit parameters in some iterative scheme. 

\begin{figure}[t]
\includegraphics[width=8cm]{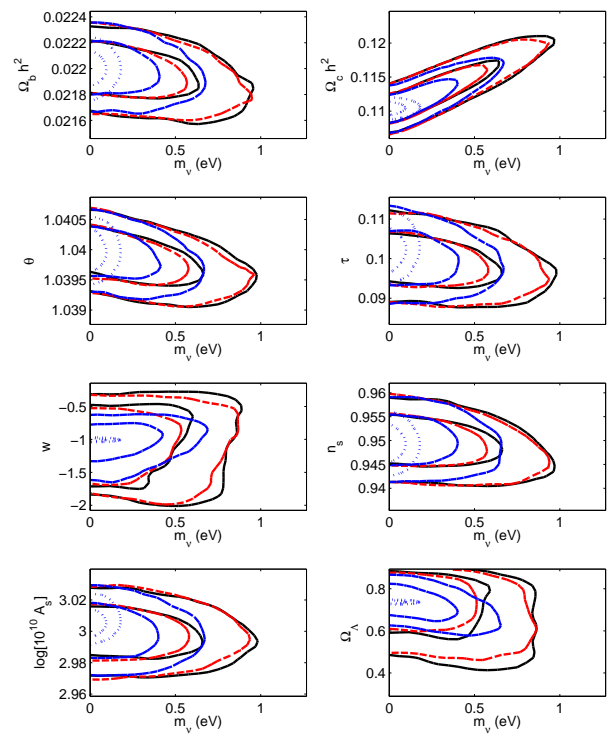}
\caption{Two-dimensional marginalized
likelihood contours involving $m_{\nu}$ obtained
by fitting some mock data mimicking the properties of Planck, DES and LSST.
The solid black curves accounts for CMB only, the red dashed for
CMB+GT with Planck+DES, the blue dashed for
CMB+GT with Planck+LSST, 
and the dotted blue for CMB+GT+GG with Planck+LSST (these combinations are
precisely defined in the text). For each case, the two lines represent
the 68\% and 95\% confidence levels.}\label{fig:mock_mnuw_2D}
\end{figure}

In this prospective, it is interesting to note that the CMB+GT
combination from Planck and LSST
increases significantly the sensitivity of Planck alone
mainly for $\Omega_{dm} h^2$ (by 30\%), $w$ (by 83\%) and $m_{\nu}$
(by 38\%). As a consequence, the sensitivity to the related parameter
$\Omega_{\Lambda}$ increases by 76\%. For our fiducial model with
$m_{\nu}=0$, the 95\% confidence level upper bound on the total
neutrino mass shrinks from 0.77~eV to 0.54~eV (for another fiducial
model with $m_{\nu}>0$ the sensitivity can only be larger than that,
see e.g. \cite{Lesgourgues:2004ps}). At this level of sensitivity, the
parameter $m_{\nu}$ is not correlated with $\Omega_{\Lambda}$ or $w$,
as can be checked by looking at two-dimensional marginalized
likelihood contours in Figure~\ref{fig:mock_mnuw_2D}. We conclude that
the cross-correlation signal derived from Planck and LSST would have
some useful sensitivity to both neutrino masses and dark energy
parameters. Instead, the correlation between Planck and DES does not bring
significant new information with respect to Planck alone. 

Ichikawa and Takahashi \cite{Ichikawa:2005hi} 
performed a similar forecast for Planck and
LSST (with slightly different specifications), using a Fisher matrix
analysis rather than MCMC approach. They find a smaller sensitivity of the
cross-correlation data to neutrino mass than we do, possibly because of
the various approximations entering into the Fisher matrix approach.

\section{Conclusions}
We have studied here the possibility to use the cross-correlation
between CMB and galaxy density maps as a tool for constraining the
neutrino mass.  On one hand massive neutrinos reduce the
cross-correlation spectrum because their free-streaming slows down
structure formation; on the other hand, they enhance it because of the
behavior of the linear growth in presence of massive neutrinos. Using
both analytic approximations and numerical computations, we showed
that in the observable range of scales and redshifts, the first effect
dominates, but the second one is not negligible. Hence the
cross-correlation between CMB and LSS maps could bring some
independent information on neutrino masses.  We performed an error
forecast analysis by fitting some mock data inspired from the Planck
satellite, Dark Energy Survey (DES) and Large Synoptic Survey
Telescope (LSST).  For Planck and LSST, the inclusion of the
cross-correlation data increases the sensitivity to $m_{\nu}$ by 38\%,
$w$ by 83\% and $\Omega_{dm} h^2$ by 30\% with respect to the CMB data
alone. With the fiducial model employed in this analysis (based on
eight free parameters) the standard deviation for the neutrino mass is
equal to 0.38~eV for Planck alone and 0.27~eV for Planck plus
cross-correlation data. This is far from being as spectacular as the
sensitivity expected from the measurement of the auto-correlation
power spectrum of future galaxy/cluster redshift surveys or cosmic
shear experiments, for which the predicted standard deviation is
closer to the level of 0.02~eV, leading to a 2$\sigma$ detection even
in the case of the minimal mass scenario allowed by current data on
neutrino oscillations (see \cite{Lesgourgues:2006nd} for a
review). However, the method proposed here is independent and affected
by different systematics. So, it remains potentially interesting, but
only if the neutrino mass is not much smaller than $m_{\nu} \sim
0.2$~eV.

\section*{Acknowledgements}
This work was initiated during a very nice and fruitful stay at the
Galileo Galilei Institute for Theoretical Physics, supported by INFN.
JL would like to thank Ofer Lohav for useful exchanges.
The project was completed thanks to the support of the EU 6th
Framework Marie Curie Research and Training network ``UniverseNet''
(MRTN-CT-2006-035863). Numerical simulations were performed on the
MUST cluster at LAPP Annecy (IN2P3/CNRS and Universit\'e de Savoie).
EG acknowledges the support from Spanish Ministerio de Ciencia y   
Tecnologia (MEC), project AYA2006-06341 with EC-FEDER funding
and research project 2005SGR00728 from  Generalitat de Catalunya.

\section*{Appendix: the Limber approximation}

Let us consider some maps $X(\hat n)$ expanded in spherical harmonics
\begin{align}
  X\left(\hat
  n\right)=\sum_{l=0}^\infty\sum_{-l}^{l}a^X_{lm}Y_{lm}(\hat n)
\end{align}
with
\begin{align}
  a^X_{lm}=\int d^2n\,Y^\ast_{lm}(\hat n)X(\hat n).\label{eq:alm_X}
\end{align}
The two-point correlation function of any two statistically
isotropic quantities $X$ and $Y$
can be expressed in terms of the power spectrum in multipole space
\begin{align}
  C_l^{XY}=\left<a^X_{lm}a^{Y\ast}_{lm}\right>,
\end{align}
or in therms of the angular
correlation function in a Legendre polynomial basis
($p_l$)
\begin{align}
  w^{XY}(\theta)=\sum_l \frac{2l+1}{4\pi}p_l(\cos\theta)C^{XY}_l.
\end{align}
In the frame of observations, a direction dependent quantity $X(\hat
n)$ is usually a quantity integrated over the line of sight, $X(\hat
n)=\int dr\,X(\vec x)$. The expression for $a^X_{lm}$,
\eqref{eq:alm_X}, can then easily be transformed to Fourier
space. Subsequently expanding the plain wave in spherical harmonics
and applying the completeness relation for spherical harmonics, one arrives at
\begin{align}
  a^X_{lm}=(-i)^l \int dr\, \frac{d^3k}{2\pi^2} X(\vec k) j_l(kr) Y^\ast_{lm}(\hat k),
\end{align}
where $X(\vec k)$ is the Fourier transform of $X(\vec x)$, $j_l(r)$ is
the spherical Bessel function, and $k=|\vec k|$. This expression can
be simplified using Limbers approximation, $\int dx\, f(x)
j_l(x) \simeq\sqrt{\frac{\pi}{2l+1}} \int dx\, f(x)
\delta(l+\tfrac12-x)$, leading to
\begin{align}
  a^X_{lm}\simeq(-i)^l\sqrt{\frac{\pi}{2l+1}} \int \frac{dr}{r} \frac{k^2 d\Omega_k}{2\pi^2} 
  X\left(\hat k, k\right) Y^\ast_{lm}(\hat k),
\end{align}
separating the $\vec k$ dependence of $X$ in $\hat k$ and $k=\frac{l+\tfrac12}{r}$.

\bibliography{refs_eg}

\begin{thebibliography}{54}
\expandafter\ifx\csname natexlab\endcsname\relax\def\natexlab#1{#1}\fi
\expandafter\ifx\csname bibnamefont\endcsname\relax
  \def\bibnamefont#1{#1}\fi
\expandafter\ifx\csname bibfnamefont\endcsname\relax
  \def\bibfnamefont#1{#1}\fi
\expandafter\ifx\csname citenamefont\endcsname\relax
  \def\citenamefont#1{#1}\fi
\expandafter\ifx\csname url\endcsname\relax
  \def\url#1{\texttt{#1}}\fi
\expandafter\ifx\csname urlprefix\endcsname\relax\def\urlprefix{URL }\fi
\providecommand{\bibinfo}[2]{#2}
\providecommand{\eprint}[2][]{\url{#2}}

\bibitem[{\citenamefont{Sachs and Wolfe}(1967)}]{Sachs:1967er}
\bibinfo{author}{\bibfnamefont{R.~K.} \bibnamefont{Sachs}} \bibnamefont{and}
  \bibinfo{author}{\bibfnamefont{A.~M.} \bibnamefont{Wolfe}},
  \bibinfo{journal}{Astrophys. J.} \textbf{\bibinfo{volume}{147}},
  \bibinfo{pages}{73} (\bibinfo{year}{1967}).

\bibitem[{\citenamefont{Crittenden and Turok}(1996)}]{Crittenden:1995ak}
\bibinfo{author}{\bibfnamefont{R.~G.} \bibnamefont{Crittenden}}
  \bibnamefont{and} \bibinfo{author}{\bibfnamefont{N.}~\bibnamefont{Turok}},
  \bibinfo{journal}{Phys. Rev. Lett.} \textbf{\bibinfo{volume}{76}},
  \bibinfo{pages}{575} (\bibinfo{year}{1996}), \eprint{astro-ph/9510072}.

\bibitem[{\citenamefont{Perlmutter et~al.}(1999)}]{Perlmutter:1998np}
\bibinfo{author}{\bibfnamefont{S.}~\bibnamefont{Perlmutter}}
  \bibnamefont{et~al.} (\bibinfo{collaboration}{Supernova Cosmology Project}),
  \bibinfo{journal}{Astrophys. J.} \textbf{\bibinfo{volume}{517}},
  \bibinfo{pages}{565} (\bibinfo{year}{1999}), \eprint{astro-ph/9812133}.

\bibitem[{\citenamefont{Riess et~al.}(1998)}]{Riess:1998cb}
\bibinfo{author}{\bibfnamefont{A.~G.} \bibnamefont{Riess}} \bibnamefont{et~al.}
  (\bibinfo{collaboration}{Supernova Search Team}), \bibinfo{journal}{Astron.
  J.} \textbf{\bibinfo{volume}{116}}, \bibinfo{pages}{1009}
  (\bibinfo{year}{1998}), \eprint{astro-ph/9805201}.

\bibitem[{\citenamefont{Spergel et~al.}(2003)}]{Spergel:2003cb}
\bibinfo{author}{\bibfnamefont{D.~N.} \bibnamefont{Spergel}}
  \bibnamefont{et~al.} (\bibinfo{collaboration}{WMAP}),
  \bibinfo{journal}{Astrophys. J. Suppl.} \textbf{\bibinfo{volume}{148}},
  \bibinfo{pages}{175} (\bibinfo{year}{2003}), \eprint{astro-ph/0302209}.

\bibitem[{\citenamefont{Spergel et~al.}(2006)}]{Spergel:2006hy}
\bibinfo{author}{\bibfnamefont{D.~N.} \bibnamefont{Spergel}}
  \bibnamefont{et~al.} (\bibinfo{collaboration}{WMAP}) (\bibinfo{year}{2006}),
  \eprint{astro-ph/0603449}.

\bibitem[{\citenamefont{Fosalba et~al.}(2003)\citenamefont{Fosalba, Gaztanaga,
  and Castander}}]{Fosalba:2003ix}
\bibinfo{author}{\bibfnamefont{P.}~\bibnamefont{Fosalba}},
  \bibinfo{author}{\bibfnamefont{E.}~\bibnamefont{Gaztanaga}},
  \bibnamefont{and}
  \bibinfo{author}{\bibfnamefont{F.}~\bibnamefont{Castander}},
  \bibinfo{journal}{ApJ} \textbf{\bibinfo{volume}{350}}, \bibinfo{pages}{L37}
  (\bibinfo{year}{2003}), \eprint{astro-ph/0305468}.

\bibitem[{\citenamefont{Boughn and Crittenden}(2004)}]{Boughn:2003yz}
\bibinfo{author}{\bibfnamefont{S.}~\bibnamefont{Boughn}} \bibnamefont{and}
  \bibinfo{author}{\bibfnamefont{R.}~\bibnamefont{Crittenden}},
  \bibinfo{journal}{Nature} \textbf{\bibinfo{volume}{427}}, \bibinfo{pages}{45}
  (\bibinfo{year}{2004}), \eprint{astro-ph/0305001}.

\bibitem[{\citenamefont{Fosalba and Gaztanaga}(2004)}]{Fosalba:2003iy}
\bibinfo{author}{\bibfnamefont{P.}~\bibnamefont{Fosalba}} \bibnamefont{and}
  \bibinfo{author}{\bibfnamefont{E.}~\bibnamefont{Gaztanaga}},
  \bibinfo{journal}{Mon. Not. Roy. Astron. Soc.}
  \textbf{\bibinfo{volume}{350}}, \bibinfo{pages}{L37} (\bibinfo{year}{2004}),
  \eprint{astro-ph/0305468}.

\bibitem[{\citenamefont{Afshordi et~al.}(2004)\citenamefont{Afshordi, Loh, and
  Strauss}}]{Afshordi:2003xu}
\bibinfo{author}{\bibfnamefont{N.}~\bibnamefont{Afshordi}},
  \bibinfo{author}{\bibfnamefont{Y.-S.} \bibnamefont{Loh}}, \bibnamefont{and}
  \bibinfo{author}{\bibfnamefont{M.~A.} \bibnamefont{Strauss}},
  \bibinfo{journal}{Phys. Rev.} \textbf{\bibinfo{volume}{D69}},
  \bibinfo{pages}{083524} (\bibinfo{year}{2004}), \eprint{astro-ph/0308260}.

\bibitem[{\citenamefont{{Padmanabhan} et~al.}(2005)\citenamefont{{Padmanabhan},
  {Hirata}, {Seljak}, {Schlegel}, {Brinkmann}, and
  {Schneider}}}]{Padmanabhan:2004}
\bibinfo{author}{\bibfnamefont{N.}~\bibnamefont{{Padmanabhan}}},
  \bibinfo{author}{\bibfnamefont{C.~M.} \bibnamefont{{Hirata}}},
  \bibinfo{author}{\bibfnamefont{U.}~\bibnamefont{{Seljak}}},
  \bibinfo{author}{\bibfnamefont{D.~J.} \bibnamefont{{Schlegel}}},
  \bibinfo{author}{\bibfnamefont{J.}~\bibnamefont{{Brinkmann}}},
  \bibnamefont{and} \bibinfo{author}{\bibfnamefont{D.~P.}
  \bibnamefont{{Schneider}}}, \bibinfo{journal}{\prd}
  \textbf{\bibinfo{volume}{72}}, \bibinfo{pages}{043525}
  (\bibinfo{year}{2005}), \eprint{arXiv:astro-ph/0410360}.

\bibitem[{\citenamefont{Cabre et~al.}(2006{\natexlab{a}})\citenamefont{Cabre,
  Gaztanaga, Manera, Fosalba, and Castander}}]{Cabre:2006qm}
\bibinfo{author}{\bibfnamefont{A.}~\bibnamefont{Cabre}},
  \bibinfo{author}{\bibfnamefont{E.}~\bibnamefont{Gaztanaga}},
  \bibinfo{author}{\bibfnamefont{M.}~\bibnamefont{Manera}},
  \bibinfo{author}{\bibfnamefont{P.}~\bibnamefont{Fosalba}}, \bibnamefont{and}
  \bibinfo{author}{\bibfnamefont{F.}~\bibnamefont{Castander}},
  \bibinfo{journal}{Mon. Not. Roy. Astron. Soc. Lett.}
  \textbf{\bibinfo{volume}{372}}, \bibinfo{pages}{L23}
  (\bibinfo{year}{2006}{\natexlab{a}}), \eprint{astro-ph/0603690}.

\bibitem[{\citenamefont{Cabre et~al.}(2006{\natexlab{b}})\citenamefont{Cabre,
  Gaztanaga, Manera, Fosalba, and Castander}}]{Cabre:2006uj}
\bibinfo{author}{\bibfnamefont{A.}~\bibnamefont{Cabre}},
  \bibinfo{author}{\bibfnamefont{E.}~\bibnamefont{Gaztanaga}},
  \bibinfo{author}{\bibfnamefont{M.}~\bibnamefont{Manera}},
  \bibinfo{author}{\bibfnamefont{P.}~\bibnamefont{Fosalba}}, \bibnamefont{and}
  \bibinfo{author}{\bibfnamefont{F.}~\bibnamefont{Castander}}
  (\bibinfo{year}{2006}{\natexlab{b}}), \eprint{astro-ph/0611046}.

\bibitem[{\citenamefont{Giannantonio et~al.}(2006)}]{Giannantonio:2006du}
\bibinfo{author}{\bibfnamefont{T.}~\bibnamefont{Giannantonio}}
  \bibnamefont{et~al.}, \bibinfo{journal}{Phys. Rev.}
  \textbf{\bibinfo{volume}{D74}}, \bibinfo{pages}{063520}
  (\bibinfo{year}{2006}), \eprint{astro-ph/0607572}.

\bibitem[{\citenamefont{{McEwen} et~al.}(2007)\citenamefont{{McEwen}, {Vielva},
  {Hobson}, {Mart{\'{\i}}nez-Gonz{\'a}lez}, and {Lasenby}}}]{McEwen:2007}
\bibinfo{author}{\bibfnamefont{J.~D.} \bibnamefont{{McEwen}}},
  \bibinfo{author}{\bibfnamefont{P.}~\bibnamefont{{Vielva}}},
  \bibinfo{author}{\bibfnamefont{M.~P.} \bibnamefont{{Hobson}}},
  \bibinfo{author}{\bibfnamefont{E.}~\bibnamefont{{Mart{\'{\i}}nez-Gonz{\'a}le%
z}}}, \bibnamefont{and} \bibinfo{author}{\bibfnamefont{A.~N.}
  \bibnamefont{{Lasenby}}}, \bibinfo{journal}{Mon. Not. Roy. Astron. Soc.
  Lett.} \textbf{\bibinfo{volume}{376}}, \bibinfo{pages}{1211}
  (\bibinfo{year}{2007}), \eprint{arXiv:astro-ph/0602398}.

\bibitem[{\citenamefont{Puchades et~al.}(2006)\citenamefont{Puchades, Fullana,
  Arnau, and Saez}}]{Puchades:2006gs}
\bibinfo{author}{\bibfnamefont{N.}~\bibnamefont{Puchades}},
  \bibinfo{author}{\bibfnamefont{M.~J.} \bibnamefont{Fullana}},
  \bibinfo{author}{\bibfnamefont{J.~V.} \bibnamefont{Arnau}}, \bibnamefont{and}
  \bibinfo{author}{\bibfnamefont{D.}~\bibnamefont{Saez}},
  \bibinfo{journal}{Mon. Not. Roy. Astron. Soc.}
  \textbf{\bibinfo{volume}{370}}, \bibinfo{pages}{1849} (\bibinfo{year}{2006}),
  \eprint{astro-ph/0605704}.

\bibitem[{\citenamefont{{Seljak} and
  {Zaldarriaga}}(2000)}]{SeljakZaldarriaga:2000}
\bibinfo{author}{\bibfnamefont{U.}~\bibnamefont{{Seljak}}} \bibnamefont{and}
  \bibinfo{author}{\bibfnamefont{M.}~\bibnamefont{{Zaldarriaga}}},
  \bibinfo{journal}{\apj} \textbf{\bibinfo{volume}{538}}, \bibinfo{pages}{57}
  (\bibinfo{year}{2000}), \eprint{arXiv:astro-ph/9907254}.

\bibitem[{\citenamefont{{Sunyaev} and {Zeldovich}}(1969)}]{SZ:1969}
\bibinfo{author}{\bibfnamefont{R.~A.} \bibnamefont{{Sunyaev}}}
  \bibnamefont{and} \bibinfo{author}{\bibfnamefont{Y.~B.}
  \bibnamefont{{Zeldovich}}}, \bibinfo{journal}{\nat}
  \textbf{\bibinfo{volume}{223}}, \bibinfo{pages}{721} (\bibinfo{year}{1969}).

\bibitem[{\citenamefont{{Afshordi} et~al.}(2007)\citenamefont{{Afshordi},
  {Lin}, {Nagai}, and {Sanderson}}}]{Afshordi:2007}
\bibinfo{author}{\bibfnamefont{N.}~\bibnamefont{{Afshordi}}},
  \bibinfo{author}{\bibfnamefont{Y.-T.} \bibnamefont{{Lin}}},
  \bibinfo{author}{\bibfnamefont{D.}~\bibnamefont{{Nagai}}}, \bibnamefont{and}
  \bibinfo{author}{\bibfnamefont{A.~J.~R.} \bibnamefont{{Sanderson}}},
  \bibinfo{journal}{Mon. Not. Roy. Astron. Soc.}
  \textbf{\bibinfo{volume}{378}}, \bibinfo{pages}{293} (\bibinfo{year}{2007}),
  \eprint{arXiv:astro-ph/0612700}.

\bibitem[{\citenamefont{Primack}(2001)}]{Primack:2001ib}
\bibinfo{author}{\bibfnamefont{J.~R.} \bibnamefont{Primack}},
  \bibinfo{journal}{SLAC Beam Line} \textbf{\bibinfo{volume}{31N3}},
  \bibinfo{pages}{50} (\bibinfo{year}{2001}), \eprint{astro-ph/0112336}.

\bibitem[{\citenamefont{Maltoni et~al.}(2004)\citenamefont{Maltoni, Schwetz,
  Tortola, and Valle}}]{Maltoni:2004ei}
\bibinfo{author}{\bibfnamefont{M.}~\bibnamefont{Maltoni}},
  \bibinfo{author}{\bibfnamefont{T.}~\bibnamefont{Schwetz}},
  \bibinfo{author}{\bibfnamefont{M.~A.} \bibnamefont{Tortola}},
  \bibnamefont{and} \bibinfo{author}{\bibfnamefont{J.~W.~F.}
  \bibnamefont{Valle}}, \bibinfo{journal}{New J. Phys.}
  \textbf{\bibinfo{volume}{6}}, \bibinfo{pages}{122} (\bibinfo{year}{2004}),
  \eprint{hep-ph/0405172}.

\bibitem[{\citenamefont{Fogli et~al.}(2006)\citenamefont{Fogli, Lisi, Marrone,
  and Palazzo}}]{Fogli:2005cq}
\bibinfo{author}{\bibfnamefont{G.~L.} \bibnamefont{Fogli}},
  \bibinfo{author}{\bibfnamefont{E.}~\bibnamefont{Lisi}},
  \bibinfo{author}{\bibfnamefont{A.}~\bibnamefont{Marrone}}, \bibnamefont{and}
  \bibinfo{author}{\bibfnamefont{A.}~\bibnamefont{Palazzo}},
  \bibinfo{journal}{Prog. Part. Nucl. Phys.} \textbf{\bibinfo{volume}{57}},
  \bibinfo{pages}{742} (\bibinfo{year}{2006}), \eprint{hep-ph/0506083}.

\bibitem[{\citenamefont{Cuoco et~al.}(2004)}]{Cuoco:2003cu}
\bibinfo{author}{\bibfnamefont{A.}~\bibnamefont{Cuoco}} \bibnamefont{et~al.},
  \bibinfo{journal}{Int. J. Mod. Phys.} \textbf{\bibinfo{volume}{A19}},
  \bibinfo{pages}{4431} (\bibinfo{year}{2004}), \eprint{astro-ph/0307213}.

\bibitem[{\citenamefont{Steigman}(2006)}]{Steigman:2005uz}
\bibinfo{author}{\bibfnamefont{G.}~\bibnamefont{Steigman}},
  \bibinfo{journal}{Int. J. Mod. Phys.} \textbf{\bibinfo{volume}{E15}},
  \bibinfo{pages}{1} (\bibinfo{year}{2006}), \eprint{astro-ph/0511534}.

\bibitem[{\citenamefont{Mangano et~al.}(2007)\citenamefont{Mangano, Melchiorri,
  Mena, Miele, and Slosar}}]{Mangano:2006ur}
\bibinfo{author}{\bibfnamefont{G.}~\bibnamefont{Mangano}},
  \bibinfo{author}{\bibfnamefont{A.}~\bibnamefont{Melchiorri}},
  \bibinfo{author}{\bibfnamefont{O.}~\bibnamefont{Mena}},
  \bibinfo{author}{\bibfnamefont{G.}~\bibnamefont{Miele}}, \bibnamefont{and}
  \bibinfo{author}{\bibfnamefont{A.}~\bibnamefont{Slosar}},
  \bibinfo{journal}{JCAP} \textbf{\bibinfo{volume}{0703}}, \bibinfo{pages}{006}
  (\bibinfo{year}{2007}), \eprint{astro-ph/0612150}.

\bibitem[{\citenamefont{Crotty et~al.}(2003)\citenamefont{Crotty, Lesgourgues,
  and Pastor}}]{Crotty:2003th}
\bibinfo{author}{\bibfnamefont{P.}~\bibnamefont{Crotty}},
  \bibinfo{author}{\bibfnamefont{J.}~\bibnamefont{Lesgourgues}},
  \bibnamefont{and} \bibinfo{author}{\bibfnamefont{S.}~\bibnamefont{Pastor}},
  \bibinfo{journal}{Phys. Rev.} \textbf{\bibinfo{volume}{D67}},
  \bibinfo{pages}{123005} (\bibinfo{year}{2003}), \eprint{astro-ph/0302337}.

\bibitem[{\citenamefont{Pierpaoli}(2003)}]{Pierpaoli:2003kw}
\bibinfo{author}{\bibfnamefont{E.}~\bibnamefont{Pierpaoli}},
  \bibinfo{journal}{Mon. Not. Roy. Astron. Soc.}
  \textbf{\bibinfo{volume}{342}}, \bibinfo{pages}{L63} (\bibinfo{year}{2003}),
  \eprint{astro-ph/0302465}.

\bibitem[{\citenamefont{Barger et~al.}(2003)\citenamefont{Barger, Kneller, Lee,
  Marfatia, and Steigman}}]{Barger:2003zg}
\bibinfo{author}{\bibfnamefont{V.}~\bibnamefont{Barger}},
  \bibinfo{author}{\bibfnamefont{J.~P.} \bibnamefont{Kneller}},
  \bibinfo{author}{\bibfnamefont{H.-S.} \bibnamefont{Lee}},
  \bibinfo{author}{\bibfnamefont{D.}~\bibnamefont{Marfatia}}, \bibnamefont{and}
  \bibinfo{author}{\bibfnamefont{G.}~\bibnamefont{Steigman}},
  \bibinfo{journal}{Phys. Lett.} \textbf{\bibinfo{volume}{B566}},
  \bibinfo{pages}{8} (\bibinfo{year}{2003}), \eprint{hep-ph/0305075}.

\bibitem[{\citenamefont{Trotta and Melchiorri}(2005)}]{Trotta:2004ty}
\bibinfo{author}{\bibfnamefont{R.}~\bibnamefont{Trotta}} \bibnamefont{and}
  \bibinfo{author}{\bibfnamefont{A.}~\bibnamefont{Melchiorri}},
  \bibinfo{journal}{Phys. Rev. Lett.} \textbf{\bibinfo{volume}{95}},
  \bibinfo{pages}{011305} (\bibinfo{year}{2005}), \eprint{astro-ph/0412066}.

\bibitem[{\citenamefont{Hannestad}(2006{\natexlab{a}})}]{Hannestad:2005jj}
\bibinfo{author}{\bibfnamefont{S.}~\bibnamefont{Hannestad}},
  \bibinfo{journal}{JCAP} \textbf{\bibinfo{volume}{0601}}, \bibinfo{pages}{001}
  (\bibinfo{year}{2006}{\natexlab{a}}), \eprint{astro-ph/0510582}.

\bibitem[{\citenamefont{Hannestad and Raffelt}(2006)}]{Hannestad:2006mi}
\bibinfo{author}{\bibfnamefont{S.}~\bibnamefont{Hannestad}} \bibnamefont{and}
  \bibinfo{author}{\bibfnamefont{G.~G.} \bibnamefont{Raffelt}},
  \bibinfo{journal}{JCAP} \textbf{\bibinfo{volume}{0611}}, \bibinfo{pages}{016}
  (\bibinfo{year}{2006}), \eprint{astro-ph/0607101}.

\bibitem[{\citenamefont{Ichikawa et~al.}(2007)\citenamefont{Ichikawa, Kawasaki,
  and Takahashi}}]{Ichikawa:2006vm}
\bibinfo{author}{\bibfnamefont{K.}~\bibnamefont{Ichikawa}},
  \bibinfo{author}{\bibfnamefont{M.}~\bibnamefont{Kawasaki}}, \bibnamefont{and}
  \bibinfo{author}{\bibfnamefont{F.}~\bibnamefont{Takahashi}},
  \bibinfo{journal}{JCAP} \textbf{\bibinfo{volume}{0705}}, \bibinfo{pages}{007}
  (\bibinfo{year}{2007}), \eprint{astro-ph/0611784}.

\bibitem[{\citenamefont{de~Bernardis et~al.}(2007)\citenamefont{de~Bernardis,
  Melchiorri, Verde, and Jimenez}}]{deBernardis:2007bu}
\bibinfo{author}{\bibfnamefont{F.}~\bibnamefont{de~Bernardis}},
  \bibinfo{author}{\bibfnamefont{A.}~\bibnamefont{Melchiorri}},
  \bibinfo{author}{\bibfnamefont{L.}~\bibnamefont{Verde}}, \bibnamefont{and}
  \bibinfo{author}{\bibfnamefont{R.}~\bibnamefont{Jimenez}}
  (\bibinfo{year}{2007}), \eprint{arXiv:0707.4170 [astro-ph]}.

\bibitem[{\citenamefont{Hamann et~al.}(2007)\citenamefont{Hamann, Hannestad,
  Raffelt, and Wong}}]{Hamann:2007pi}
\bibinfo{author}{\bibfnamefont{J.}~\bibnamefont{Hamann}},
  \bibinfo{author}{\bibfnamefont{S.}~\bibnamefont{Hannestad}},
  \bibinfo{author}{\bibfnamefont{G.~G.} \bibnamefont{Raffelt}},
  \bibnamefont{and} \bibinfo{author}{\bibfnamefont{Y.~Y.~Y.}
  \bibnamefont{Wong}}, \bibinfo{journal}{JCAP} \textbf{\bibinfo{volume}{0708}},
  \bibinfo{pages}{021} (\bibinfo{year}{2007}), \eprint{arXiv:0705.0440
  [astro-ph]}.

\bibitem[{\citenamefont{Bond et~al.}(1980)\citenamefont{Bond, Efstathiou, and
  Silk}}]{Bond:1980ha}
\bibinfo{author}{\bibfnamefont{J.~R.} \bibnamefont{Bond}},
  \bibinfo{author}{\bibfnamefont{G.}~\bibnamefont{Efstathiou}},
  \bibnamefont{and} \bibinfo{author}{\bibfnamefont{J.}~\bibnamefont{Silk}},
  \bibinfo{journal}{Phys. Rev. Lett.} \textbf{\bibinfo{volume}{45}},
  \bibinfo{pages}{1980} (\bibinfo{year}{1980}).

\bibitem[{\citenamefont{Hannestad}(2007)}]{Hannestad:2007tu}
\bibinfo{author}{\bibfnamefont{S.}~\bibnamefont{Hannestad}}
  (\bibinfo{year}{2007}), \eprint{arXiv:0710.1952 [hep-ph]}.

\bibitem[{\citenamefont{Hannestad}(2006{\natexlab{b}})}]{Hannestad:2006zg}
\bibinfo{author}{\bibfnamefont{S.}~\bibnamefont{Hannestad}},
  \bibinfo{journal}{Ann. Rev. Nucl. Part. Sci.} \textbf{\bibinfo{volume}{56}},
  \bibinfo{pages}{137} (\bibinfo{year}{2006}{\natexlab{b}}),
  \eprint{hep-ph/0602058}.

\bibitem[{\citenamefont{Lesgourgues and Pastor}(2006)}]{Lesgourgues:2006nd}
\bibinfo{author}{\bibfnamefont{J.}~\bibnamefont{Lesgourgues}} \bibnamefont{and}
  \bibinfo{author}{\bibfnamefont{S.}~\bibnamefont{Pastor}},
  \bibinfo{journal}{Phys. Rept.} \textbf{\bibinfo{volume}{429}},
  \bibinfo{pages}{307} (\bibinfo{year}{2006}), \eprint{astro-ph/0603494}.

\bibitem[{\citenamefont{Lesgourgues et~al.}(2004)\citenamefont{Lesgourgues,
  Pastor, and Perotto}}]{Lesgourgues:2004ps}
\bibinfo{author}{\bibfnamefont{J.}~\bibnamefont{Lesgourgues}},
  \bibinfo{author}{\bibfnamefont{S.}~\bibnamefont{Pastor}}, \bibnamefont{and}
  \bibinfo{author}{\bibfnamefont{L.}~\bibnamefont{Perotto}},
  \bibinfo{journal}{Phys. Rev.} \textbf{\bibinfo{volume}{D70}},
  \bibinfo{pages}{045016} (\bibinfo{year}{2004}), \eprint{hep-ph/0403296}.

\bibitem[{\citenamefont{Wang et~al.}(2005)\citenamefont{Wang, Haiman, Hu,
  Khoury, and May}}]{Wang:2005vr}
\bibinfo{author}{\bibfnamefont{S.}~\bibnamefont{Wang}},
  \bibinfo{author}{\bibfnamefont{Z.}~\bibnamefont{Haiman}},
  \bibinfo{author}{\bibfnamefont{W.}~\bibnamefont{Hu}},
  \bibinfo{author}{\bibfnamefont{J.}~\bibnamefont{Khoury}}, \bibnamefont{and}
  \bibinfo{author}{\bibfnamefont{M.}~\bibnamefont{May}},
  \bibinfo{journal}{Phys. Rev. Lett.} \textbf{\bibinfo{volume}{95}},
  \bibinfo{pages}{011302} (\bibinfo{year}{2005}), \eprint{astro-ph/0505390}.

\bibitem[{\citenamefont{Hannestad and Wong}(2007)}]{Hannestad:2007cp}
\bibinfo{author}{\bibfnamefont{S.}~\bibnamefont{Hannestad}} \bibnamefont{and}
  \bibinfo{author}{\bibfnamefont{Y.~Y.~Y.} \bibnamefont{Wong}},
  \bibinfo{journal}{JCAP} \textbf{\bibinfo{volume}{0707}}, \bibinfo{pages}{004}
  (\bibinfo{year}{2007}), \eprint{astro-ph/0703031}.

\bibitem[{\citenamefont{Song and Knox}(2003)}]{Song:2003gg}
\bibinfo{author}{\bibfnamefont{Y.-S.} \bibnamefont{Song}} \bibnamefont{and}
  \bibinfo{author}{\bibfnamefont{L.}~\bibnamefont{Knox}}
  (\bibinfo{year}{2003}), \eprint{astro-ph/0312175}.

\bibitem[{\citenamefont{Hannestad et~al.}(2006)\citenamefont{Hannestad, Tu, and
  Wong}}]{Hannestad:2006as}
\bibinfo{author}{\bibfnamefont{S.}~\bibnamefont{Hannestad}},
  \bibinfo{author}{\bibfnamefont{H.}~\bibnamefont{Tu}}, \bibnamefont{and}
  \bibinfo{author}{\bibfnamefont{Y.~Y.~Y.} \bibnamefont{Wong}},
  \bibinfo{journal}{JCAP} \textbf{\bibinfo{volume}{0606}}, \bibinfo{pages}{025}
  (\bibinfo{year}{2006}), \eprint{astro-ph/0603019}.

\bibitem[{\citenamefont{Ichikawa and Takahashi}(2005)}]{Ichikawa:2005hi}
\bibinfo{author}{\bibfnamefont{K.}~\bibnamefont{Ichikawa}} \bibnamefont{and}
  \bibinfo{author}{\bibfnamefont{T.}~\bibnamefont{Takahashi}}
  (\bibinfo{year}{2005}), \eprint{astro-ph/0510849}.

\bibitem[{\citenamefont{Kiakotou et~al.}(2007)\citenamefont{Kiakotou, Elgaroy,
  and Lahav}}]{Kiakotou:2007pz}
\bibinfo{author}{\bibfnamefont{A.}~\bibnamefont{Kiakotou}},
  \bibinfo{author}{\bibfnamefont{O.}~\bibnamefont{Elgaroy}}, \bibnamefont{and}
  \bibinfo{author}{\bibfnamefont{O.}~\bibnamefont{Lahav}}
  (\bibinfo{year}{2007}), \eprint{arXiv:0709.0253 [astro-ph]}.

\bibitem[{\citenamefont{Smith et~al.}(2003)}]{Smith:2002dz}
\bibinfo{author}{\bibfnamefont{R.~E.} \bibnamefont{Smith}} \bibnamefont{et~al.}
  (\bibinfo{collaboration}{The Virgo Consortium}), \bibinfo{journal}{Mon. Not.
  Roy. Astron. Soc.} \textbf{\bibinfo{volume}{341}}, \bibinfo{pages}{1311}
  (\bibinfo{year}{2003}), \eprint{astro-ph/0207664}.

\bibitem[{\citenamefont{{Cabre} et~al.}(2007)\citenamefont{{Cabre}, {Fosalba},
  {Gaztanaga}, and {Manera}}}]{Cabre.etal:2007}
\bibinfo{author}{\bibfnamefont{A.}~\bibnamefont{{Cabre}}},
  \bibinfo{author}{\bibfnamefont{P.}~\bibnamefont{{Fosalba}}},
  \bibinfo{author}{\bibfnamefont{E.}~\bibnamefont{{Gaztanaga}}},
  \bibnamefont{and} \bibinfo{author}{\bibfnamefont{M.}~\bibnamefont{{Manera}}},
  \bibinfo{journal}{ArXiv Astrophysics e-prints}  (\bibinfo{year}{2007}),
  \eprint{astro-ph/0701393}.

\bibitem[{\citenamefont{Perotto et~al.}(2006)\citenamefont{Perotto,
  Lesgourgues, Hannestad, Tu, and Wong}}]{Perotto:2006rj}
\bibinfo{author}{\bibfnamefont{L.}~\bibnamefont{Perotto}},
  \bibinfo{author}{\bibfnamefont{J.}~\bibnamefont{Lesgourgues}},
  \bibinfo{author}{\bibfnamefont{S.}~\bibnamefont{Hannestad}},
  \bibinfo{author}{\bibfnamefont{H.}~\bibnamefont{Tu}}, \bibnamefont{and}
  \bibinfo{author}{\bibfnamefont{Y.~Y.~Y.} \bibnamefont{Wong}},
  \bibinfo{journal}{JCAP} \textbf{\bibinfo{volume}{0610}}, \bibinfo{pages}{013}
  (\bibinfo{year}{2006}), \eprint{astro-ph/0606227}.

\bibitem[{\citenamefont{Lesgourgues et~al.}(2006)\citenamefont{Lesgourgues,
  Perotto, Pastor, and Piat}}]{Lesgourgues:2005yv}
\bibinfo{author}{\bibfnamefont{J.}~\bibnamefont{Lesgourgues}},
  \bibinfo{author}{\bibfnamefont{L.}~\bibnamefont{Perotto}},
  \bibinfo{author}{\bibfnamefont{S.}~\bibnamefont{Pastor}}, \bibnamefont{and}
  \bibinfo{author}{\bibfnamefont{M.}~\bibnamefont{Piat}},
  \bibinfo{journal}{Phys. Rev.} \textbf{\bibinfo{volume}{D73}},
  \bibinfo{pages}{045021} (\bibinfo{year}{2006}), \eprint{astro-ph/0511735}.

\bibitem[{\citenamefont{Pogosian et~al.}(2005)\citenamefont{Pogosian,
  Corasaniti, Stephan-Otto, Crittenden, and Nichol}}]{Pogosian:2005ez}
\bibinfo{author}{\bibfnamefont{L.}~\bibnamefont{Pogosian}},
  \bibinfo{author}{\bibfnamefont{P.~S.} \bibnamefont{Corasaniti}},
  \bibinfo{author}{\bibfnamefont{C.}~\bibnamefont{Stephan-Otto}},
  \bibinfo{author}{\bibfnamefont{R.}~\bibnamefont{Crittenden}},
  \bibnamefont{and} \bibinfo{author}{\bibfnamefont{R.}~\bibnamefont{Nichol}},
  \bibinfo{journal}{Phys. Rev.} \textbf{\bibinfo{volume}{D72}},
  \bibinfo{pages}{103519} (\bibinfo{year}{2005}), \eprint{astro-ph/0506396}.

\bibitem[{\citenamefont{LoVerde et~al.}(2007)\citenamefont{LoVerde, Hui, and
  Gaztanaga}}]{LoVerde:2006cj}
\bibinfo{author}{\bibfnamefont{M.}~\bibnamefont{LoVerde}},
  \bibinfo{author}{\bibfnamefont{L.}~\bibnamefont{Hui}}, \bibnamefont{and}
  \bibinfo{author}{\bibfnamefont{E.}~\bibnamefont{Gaztanaga}},
  \bibinfo{journal}{Phys. Rev.} \textbf{\bibinfo{volume}{D75}},
  \bibinfo{pages}{043519} (\bibinfo{year}{2007}), \eprint{astro-ph/0611539}.

\bibitem[{\citenamefont{Lewis and Bridle}(2002)}]{Lewis:2002ah}
\bibinfo{author}{\bibfnamefont{A.}~\bibnamefont{Lewis}} \bibnamefont{and}
  \bibinfo{author}{\bibfnamefont{S.}~\bibnamefont{Bridle}},
  \bibinfo{journal}{Phys. Rev.} \textbf{\bibinfo{volume}{D66}},
  \bibinfo{pages}{103511} (\bibinfo{year}{2002}), \eprint{astro-ph/0205436}.

\bibitem[{\citenamefont{Doran and Robbers}(2006)}]{Doran:2006kp}
\bibinfo{author}{\bibfnamefont{M.}~\bibnamefont{Doran}} \bibnamefont{and}
  \bibinfo{author}{\bibfnamefont{G.}~\bibnamefont{Robbers}},
  \bibinfo{journal}{JCAP} \textbf{\bibinfo{volume}{0606}}, \bibinfo{pages}{026}
  (\bibinfo{year}{2006}), \eprint{astro-ph/0601544}.

\bibitem[{\citenamefont{Doran et~al.}(2007)\citenamefont{Doran, Robbers, and
  Wetterich}}]{Doran:2005ep}
\bibinfo{author}{\bibfnamefont{M.}~\bibnamefont{Doran}},
  \bibinfo{author}{\bibfnamefont{G.}~\bibnamefont{Robbers}}, \bibnamefont{and}
  \bibinfo{author}{\bibfnamefont{C.}~\bibnamefont{Wetterich}},
  \bibinfo{journal}{Phys. Rev.} \textbf{\bibinfo{volume}{D75}},
  \bibinfo{pages}{023003} (\bibinfo{year}{2007}), \eprint{astro-ph/0609814}.

\end{thebibliography}

\end{document}